\documentclass[twocolumn,prd,floatfix,preprintnumbers,a4paper,nofootinbib,superscriptaddress]{revtex4-2}
\usepackage[utf8]{inputenc}
\usepackage{amsmath,amssymb}
\usepackage{amsfonts}
\usepackage{bm}
\usepackage{courier}
\usepackage{graphics}
\usepackage{graphicx}
\usepackage{comment}
\usepackage{float}
\usepackage{amsmath}
\usepackage{amssymb}
\usepackage[normalem]{ulem} 
\usepackage[mathscr]{euscript}
\usepackage{acronym}
\usepackage{float}
\usepackage[caption=false]{subfig}
\usepackage{multirow}
\usepackage{graphicx}
\usepackage{multirow}
\usepackage{graphicx}
\usepackage[usenames,dvipsnames,table,xcdraw]{xcolor}
\usepackage[colorlinks, pdfborder={0 0 0}]{hyperref} 
\usepackage[nameinlink,capitalise]{cleveref}
\usepackage{tensor}
\usepackage{verbatim}
\usepackage{wrapfig}
\usepackage{soul} 

\graphicspath{{fig}}

\newcommand{\sapienza}{Dipartimento di Fisica, Sapienza Università 
	di Roma, Piazzale Aldo Moro 5, 00185, Roma, Italy}
\newcommand{\infn}{INFN, Sezione di Roma, Piazzale Aldo Moro 2, 00185, Roma, Italy}
\newcommand{\uib}{
Departament de Física, Universitat de les Illes Balears, IAC3 – IEEC, Crta. Valldemossa km 7.5,
E-07122 Palma, Spain}

\begin{document}
\title{Black Hole Ringdown Amplitudescopy}

\author{Francesco Crescimbeni}
\email{francesco.crescimbeni@uniroma1.it}
\affiliation{\sapienza}
\affiliation{\infn}

\author{Xisco Jimenez Forteza}
\email{f.jimenez@uib.es}
\affiliation{\uib}

\author{Paolo Pani}
\email{paolo.pani@uniroma1.it}
\affiliation{\sapienza}
\affiliation{\infn}

\begin{abstract}
Black hole ringdowns in extensions of General Relativity (GR) generically exhibit two distinct signatures: i) theory-dependent shifts in the standard black-hole quasinormal modes, and ii) additional modes arising from extra fundamental fields --~such as scalar, vector, or tensor degrees of freedom~-- that can also contribute to the gravitational-wave signal.
As recently argued, in general \emph{both} effects are present simultaneously, and accurately modeling them is essential for robust tests of GR in the ringdown regime. 
In this work, we investigate the impact of extra field-induced modes, which are often neglected in standard ringdown analyses, on the interpretation of gravitational-wave signals. 
To provide some concrete examples, we focus on dynamical Chern-Simons and Einstein-scalar-Gauss-Bonnet theories, 
well-motivated extensions of GR, characterized respectively by a parity-odd and a parity-even coupling between a dynamical scalar field and quadratic curvature invariants. 
We show that including extra field-induced modes improves the bounds on these theories compared to standard spectroscopy and also allows for equally constraining complementary tests not based on quasinormal mode shifts.
Our analysis highlights the relevance of incorporating extra field-induced modes in ringdown templates and assesses their potential to either bias or enhance constraints on GR deviations.
\end{abstract}

\preprint{ET-0444A-25.}

\maketitle

\section{Introduction}
%
The black hole (BH) spectroscopy program~\cite{Dreyer:2003bv,Detweiler:1980gk,Berti:2005ys,Gossan:2011ha} is a central component of strong-field tests of General Relativity (GR)~\cite{LIGOScientific:2021sio,Berti:2015itd,Berti:2018vdi,Colleoni:2024lpj}, offering a unique avenue for probing the nature of compact objects formed in the aftermath of binary coalescences~\cite{Cardoso:2019rvt} (see~\cite{Berti:2025hly} for a recent review). This approach focuses on analyzing the ringdown phase of gravitational-wave (GW) signals, where the newly formed remnant emits radiation that, in a certain time frame, can be characterized by a superposition of quasinormal modes (QNMs)~\cite{Vishveshwara:1970zz,Kokkotas:1999bd,Berti:2009kk,Konoplya:2011qq}.

Within the framework of linear perturbation theory, the GW signal $h(t)$ at intermediate times after merger is modeled as a sum of damped sinusoids corresponding to the remnant’s QNMs~\cite{Leaver:1986gd}, which reflects the BH relaxation toward a stationary configuration. 
If the remnant is a BH, GR predicts that its full QNM spectrum is completely determined by only two parameters: the remnant's mass $M$ and dimensionless spin $\chi$. This distinctive property enables stringent null tests of GR~\cite{Brito:2018rfr,Carullo:2019flw,Isi:2019aib,LIGOScientific:2021sio,Forteza:2022tgq,Franchini:2023eda,Ma:2023cwe,Ma:2022wpv,Baibhav:2023clw} and serves as a robust tool for probing the fundamental nature of the remnant~\cite{Maggio:2020jml,Maggio:2021ans,Maggio:2023fwy}. Very recently, the LIGO-Virgo-KAGRA~(LVK) Collaboration performed the first robust spectroscopy test using the exceptional GW250114 event~\cite{LIGOScientific:2025epi,LIGOScientific:2025obp}, constraining the first overtone frequency with an accuracy of approximately $30\%$, and finding some evidence for other modes in the ringdown of this event. 

Almost any theory beyond GR predicts extra degrees of freedom coupled to gravity~\cite{Sotiriou:2014yhm,Berti:2015itd}.
These extra fields (whether scalars~\cite{Horndeski:1974wa,Kanti:1995vq,Alexander:2009tp,Endlich:2017tqa}, vectors~\cite{Jacobson:2000xp,Horava:2009uw,Heisenberg:2014rta}, tensors~\cite{deRham:2010kj,Hassan:2011zd,Antoniou:2024jku}, etc, depending on the theory) may modify the stationary BH solutions, leading to deviations from the Kerr metric, and/or modify the dynamics of the theory. As recently put forward~\cite{Crescimbeni:2024sam}, in either case two generic predictions are: i) a deformation of the standard Kerr QNMs,
and ii) the existence of extra field-induced modes in the gravitational signal, that can be excited during the ringdown~\cite{Molina:2010fb,Pani:2013pma,Blazquez-Salcedo:2016enn,Cardoso:2020nst,Antoniou:2024jku,Crescimbeni:2024sam,Lestingi:2025jyb}. 

In this work we build on the recent analysis of~\cite{Crescimbeni:2024sam} by applying their framework to two specific examples: dynamical Chern-Simons theory~\cite{Alexander:2009tp} and Einstein-scalar-Gauss-Bonnet theory~\cite{Kanti:1995vq}. Both theories extend GR by introducing a (parity-odd and parity-even, respectively) scalar field, coupled to quadratic curvature invariants. Indeed, these theories are among the best studied high-curvature deviations from GR and serve as complementary representative examples for quadratic curvature corrections which might be relevant during the merger of compact objects.
Furthermore, the BH QNM spectrum and the excitation of the extra scalar modes in GW signals in these theories have been studied in some detail~\cite{Pani:2009wy,Cardoso:2009pk,Molina:2010fb,Pani:2013ija,Pani:2013wsa,Pierini:2021jxd,Wagle:2021tam,Cano:2021myl,Pierini:2022eim,Cano:2023jbk,Wagle:2023fwl,Chung:2023zdq,Chung:2023wkd,Blazquez-Salcedo:2023hwg,Chung:2024ira,Chung:2024vaf}, making them two perfect testbeds to quantify the impact of extra modes in the ringdown waveform, which is our main goal in this paper.
Henceforth we will use $G=c=1$ units.
\section{Framework}
As recently proposed in~\cite{Crescimbeni:2024sam}, a general model for the ringdown waveform in theories beyond GR can be written as \footnote{For simplicity, we assume circular polarization.}
\begin{align}\label{eq:rdmodel2}
    h(t)=&\sum_{\ell mn} A_{\ell mn}\cos\left(2\pi f_{\ell mn}^{\rm grav}t+\phi_{\ell mn}\right)
    e^{-{t}/{\tau_{\ell mn}^{\rm grav}}}\nonumber\\
    &+\sum_{\ell mn} \hat A_{\ell mn}\cos\left(2\pi \hat f_{\ell mn}t+\hat \phi_{\ell mn}\right)e^{-{t}/{\hat\tau_{\ell mn}}}\,,
\end{align}
where $A_{\ell mn}$, $\phi_{\ell mn}$, $f_{\ell mn}$, and $\tau_{\ell mn}$ respectively denote the amplitude, phase, frequency, and damping time of the $(\ell,m,n)$ mode, with $(\ell,m,n)$ denoting the angular, azimuthal, and overtone number, respectively. The fundamental mode is denoted by $n=0$, with $n>0$ tones having shorter damping time.
We shall use the hat to denote quantities related to the extra (scalar, vector, tensor, etc) field-induced modes, whereas $f_i^{\rm grav}$ and $\tau_i^{\rm grav}$ are the frequency and damping time of the $i$-th gravitational mode, where the index $i=(\ell mn)$ is a short-hand notation.

We will focus on small deviations to GR, which are motivated by considering Einstein's theory as the lowest term in an effective-field-theory expansion, and also by the stringent observational constraints already placed by GW observations~\cite{LIGOScientific:2021sio,Berti:2025hly,LIGOScientific:2025epi,LIGOScientific:2025obp}. We can therefore write the QNMs as\footnote{For concreteness, we shall assume that the extra modes are scalar degrees of freedom, but our formalism is general and applies straightforwardly to any other cases.}
\begin{align}
    f_i^{\rm grav}=f_i^{\rm Kerr}(1+\delta f_i)\,,&\quad \tau_i^{\rm grav}=\tau_i^{\rm Kerr}(1+\delta \tau_i)\,, \label{eq:QNMmod}\\
    \hat f_i=f_i^{{\rm Kerr},\,s=0}(1+\delta \hat f_i)\,,&\quad \hat \tau_i=\tau_i^{{\rm Kerr},\,s=0}(1+\delta \hat \tau_i)\,. \label{eq:QNMmod2}
\end{align}
In the above expression, $f_i^{\rm Kerr}$ and $\tau_i^{\rm Kerr}$ (resp., $f_i^{{\rm Kerr},\,s=0}$ and $\tau_i^{{\rm Kerr},\,s=0}$) are the standard gravitational (resp., scalar) QNM frequency and damping time of a Kerr 
BH in GR, and depend only on the remnant mass $M$ and spin $\chi$.
The quantities
$\delta X$'s collectively denote small, theory-dependent deviations from the leading-order quantity $X$.
These corrections depend generically on $(M,\chi)$, but also on any fundamental coupling constant of a given GR extension. Their explicit expression must be compute on a case-by-case basis, either perturbatively in the spin~\cite{Pani:2009wy,Cardoso:2009pk,Molina:2010fb,Pani:2013ija,Pani:2013wsa,Pierini:2021jxd,Wagle:2021tam,Cano:2021myl,Pierini:2022eim,Cano:2023jbk,Wagle:2023fwl} or fully numerically~\cite{Dias:2015wqa,Chung:2023zdq,Chung:2023wkd,Blazquez-Salcedo:2023hwg,Chung:2024ira,Chung:2024vaf}.
An alternative approach is to parametrize the QNM deviations, either assuming that $\delta X$'s are constant (independent of the BH parameters and theory couplings)~\cite{LIGOScientific:2021sio,Ma:2023cwe,Ma:2022wpv} or, more realistically, by considering a small-spin expansion of each deviation~\cite{Maselli:2019mjd,Carullo:2021dui,Maselli:2023khq}, which inevitably inflates the number of free parameters in the model.

However, when it comes to the extra field-induced modes, an important simplification occurs~\cite{Crescimbeni:2024sam}:
since the amplitudes $\hat A_i$ of these extra modes are \emph{proportional} to (powers of) the coupling constants and hence vanish in the GR limit~\cite{Cardoso:2009pk,Molina:2010fb,Blazquez-Salcedo:2016enn,Cardoso:2020nst}, to leading order in the corrections one can neglect $\delta \hat f_i$ and $\delta \hat \tau_i$, so that the GR deviations are generically parametrized only by the amplitude of the \emph{test-field} modes, independently of the theory (see, e.g.,~\cite{BertiWebpage,CardosoWebpage} for tabulated values of test-field QNMs in Kerr).
Since these corrections are generic and involve only amplitudes and phases rather than frequencies, we dub a test based on extra field-induced modes as \emph{BH ringdown amplitudescopy}, to distinguish it from the usual BH ringdown spectroscopy which involves measuring shifts in the QNM frequencies.

Overall, to leading order in the GR deviations, the most general ringdown waveform can be written as:
\begin{align}\label{eq:rdmodelFINAL}
    h(t)=&\sum_{i} A_{i}\cos\left(2\pi f_i^{\rm Kerr}(1+\delta f_i)t+\phi_{i}\right)
    e^{-\frac{t}{\tau_i^{\rm Kerr}(1+\delta\tau_i)}}\nonumber\\
    &+\sum_{i} \hat A_{i}\cos\left(2\pi \hat f_{i}^{{\rm Kerr},\,s=0}t+\hat \phi_{i}\right)e^{-{t}/{\hat\tau_{i}^{{\rm Kerr},\,s=0}}}\,.
\end{align}
Ordinary ringdown tests of gravity are based on the first line of the above waveform, whereas the extra field-induced modes enter in the second line, remarkably in a model-agnostic fashion, since the only beyond-GR parameters are the amplitudes $\hat A_i$ and phases $\hat \phi_i$~\cite{Crescimbeni:2024sam}. Very recently, this approach has been formalized in~\cite{Lestingi:2025jyb}, showing that Eq.~\eqref{eq:rdmodelFINAL}, derived in~\cite{Crescimbeni:2024sam}, is the most general ringdown waveform in the context of small GR deviations.


\section{Examples: Chern-Simons and Gauss-Bonnet gravity theories}
%
Our case studies will involve dynamical Chern-Simons (henceforth, CS) and Einstein-scalar-Gauss-Bonnet (henceforth, GB) gravity theories. 
CS theory is defined by the action (following the conventions in Ref.~\cite{Perkins:2021mhb})
%
\begin{eqnarray}
&&S_{\rm CS}=\int d^4x\sqrt{-g}R-\frac{1}{2}\int d^4x\sqrt{-g}
g^{\mu\nu}\nabla_\mu\vartheta\nabla_\nu\vartheta\nonumber\\
&&+\alpha_{\rm CS}\int d^4x\sqrt{-g}
\vartheta\,^*RR\,.\label{actionCS}
\end{eqnarray}
where $\vartheta$ is a (pseudo-)scalar field, $^*RR=\frac{1}{2}R_{\alpha\beta\gamma\delta}\epsilon^{\beta\alpha \epsilon\phi}R^{\gamma\delta}_{~~\epsilon\phi}$ is an odd-parity quadratic-curvature invariant, and $\alpha_{\rm CS}\equiv\ell^2_{\rm CS}$ is the coupling constant.

GB gravity is instead defined as (following the conventions in Refs.~\cite{Chung:2024vaf,Chung:2025gyg,Chung:2025wbg})
\begin{eqnarray}
&&S_{\rm GB}=\int d^4x\sqrt{-g}R-\frac{1}{2}\int d^4x\sqrt{-g}
g^{\mu\nu}\nabla_\mu\vartheta\nabla_\nu\vartheta\nonumber\\
&&+\alpha_{\rm GB}\int d^4x\sqrt{-g}
\vartheta\,(R^2-4 R_{\mu\nu}R^{\mu\nu}+R_{\alpha\beta\gamma\delta}R^{\alpha\beta\gamma\delta})\,.\nonumber\\\label{actionGB}
\end{eqnarray}
In this case $\vartheta$ is an ordinary scalar field (the dilaton, when the theory is framed as a low-energy truncation of string theory) which couples to the (even-parity) Gauss-Bonnet topological invariant. Since also in this theory the coupling has the dimensions of a length squared, we define $\alpha_{\rm GB}\equiv\ell_{\rm GB}^2$.

\begin{figure*}[!t]
    \centering
    \includegraphics[width=0.9\linewidth]{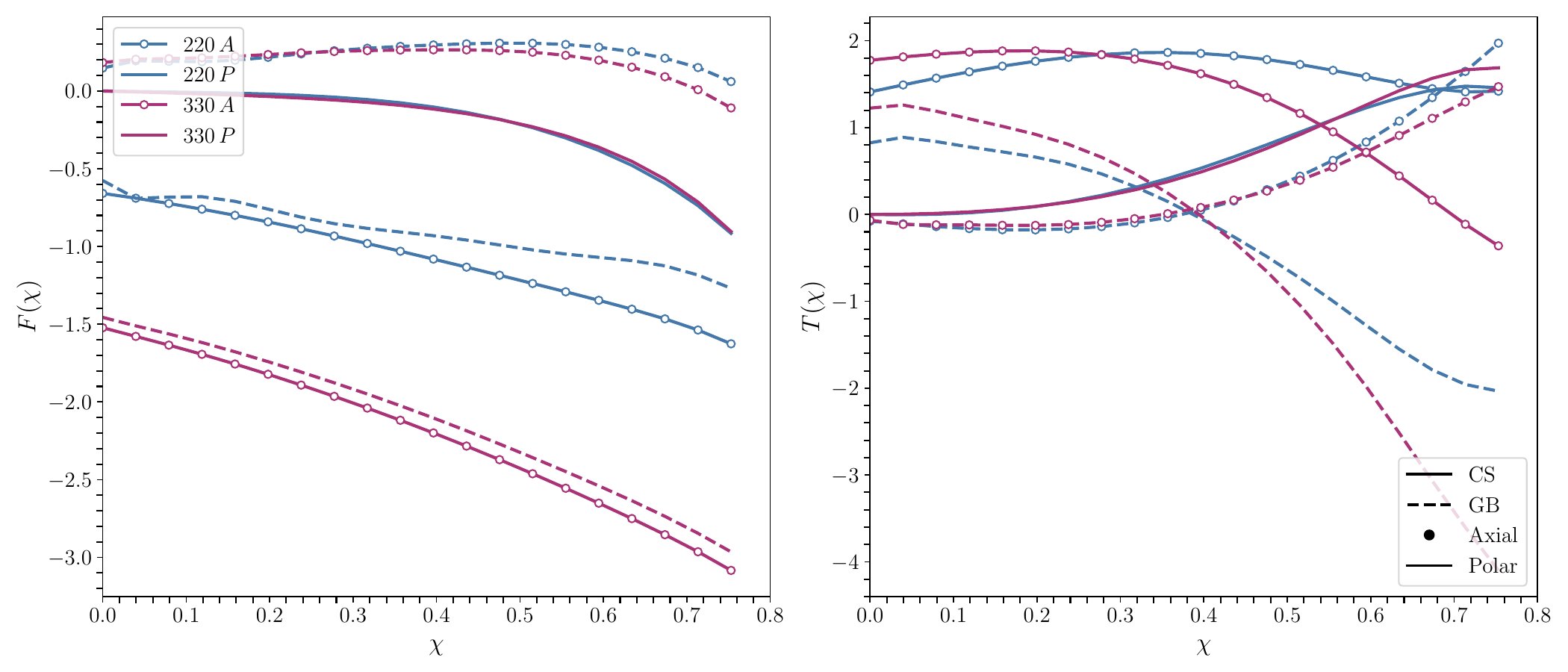}
    \caption{Fits for the corrections to the QNM frequency and damping time as a function of the spin $\chi$ for CS gravity~\cite{Chung:2025gyg} 
    (continuous lines) and GB gravity~\cite{Chung:2024vaf} (dashed lines). For each theory, we show the deviations of the $(220)$ (blue) and $(330)$ (purple) modes. Dotted and continuous curves refer to axial and polar modes, respectively.}
    \label{fig:dCS_deviations}
\end{figure*}

While spherically-symmetric BH solutions to CS gravity are equivalent to the GR Schwarzschild solution, in both theories spinning BHs are different from Kerr and are endowed with a nontrivial scalar field. Furthermore, the field equations couple perturbations of the background scalar field to those of the metric~\cite{Molina:2010fb,Pani:2013pma,Blazquez-Salcedo:2016enn,Chung:2024ira,Chung:2024vaf,Lestingi:2025jyb}, resulting in different effective potentials and coupling terms, even in the non-rotating case.

The QNMs of spinning BHs in these theories have been originally compute in a small-spin expansion~\cite{Pani:2009wy,Cardoso:2009pk,Molina:2010fb,Pani:2013ija,Pani:2013wsa,Pierini:2021jxd,Wagle:2021tam,Cano:2021myl,Pierini:2022eim,Cano:2023jbk,Wagle:2023fwl} and, more recently, fully numerically~\cite{Chung:2023zdq,Chung:2023wkd,Blazquez-Salcedo:2023hwg,Chung:2024ira,Chung:2024vaf}.
Here we will make use of the fitting formulas for the gravitational modes provided in Table~V of Ref.~\cite{Chung:2024vaf} for GB gravity, and Table~V of Ref.~\cite{Chung:2025gyg} for CS theory.
These fits provide
\begin{eqnarray}
    \delta f_i &= \zeta F_i(\chi)\,,\qquad
    \delta \tau_i &= -\frac{\zeta T_i(\chi)}{1+\zeta T_i(\chi)}\,,
    \label{corrections}
\end{eqnarray}
where $\zeta T_i$ is the deviation of the imaginary part of the $i$-th QNM, namely the inverse of the damping time $\tau_i$. In Appendix~\ref{sec:fs_computation}, we review a derivation of Eq.~\eqref{corrections} and provide the fitting coefficients. The dependence on the coupling constant and remnant mass can be factored out into a dimensionless coupling $\zeta=\frac{\alpha^2_{\rm CS/GB}}{M^4}$, so that the functions $F_i$ and $T_i$ (for any given $i$-th mode) depend only on the remnant spin, and are of course different for the two theories. 
In Fig.~\ref{fig:dCS_deviations}, we plot $F_i$ and $T_i$ for the most relevant QNMs in these two theories. 
Note that beyond-GR effects break the isospectrality~\cite{Chandrasekhar:1985kt} of the gravitational modes that can be classified as polar and axial in the non-spinning case~\cite{Pani:2013ija,Barausse:2014tra,Li:2023ulk}. As a result, each $(\ell m n)$ mode is further split into an axial~(A) and polar~(P) mode. As a check, we can observe that, in the zero-spin limit for CS gravity, the deviations of the polar frequencies and damping times approach zero~\cite{Molina:2010fb}.

Finally, in both theories, the extra modes will be those of a \emph{test} scalar field in Kerr, which are tabulated as a function of the mass and spin~\cite{BertiWebpage,CardosoWebpage}. To the leading order, also the amplitudes of these modes scale with the coupling constant~\cite{Molina:2010fb,Pani:2013pma,Blazquez-Salcedo:2016enn,Chung:2024ira,Chung:2024vaf,Lestingi:2025jyb}
\begin{equation}
    \hat A_{\ell m n} = \gamma_{\ell m n} \zeta\,.
\end{equation}
Putative higher-order corrections are consistently neglected within our framework.
For a given theory,
the normalized amplitudes $\gamma_{\ell m n}$ and the phases $\hat \phi_{\ell m n}$ depend on the properties of the progenitor binary. Here we will just consider them as free parameters of the model. Since we will only consider $(220)$ scalar modes, we will simplify the notation by defining $\gamma\equiv \gamma_{220}^{s=0}$.

Note that the perturbative approach imposes $\hat A_{\ell m n}\ll A_{\ell m n}$, but $\gamma$ can be larger than unit as long as $\zeta\ll1$. For example, for typical values adopted below ($\ell_{\rm GB/CS}\approx 35\,{\rm km}$, $M\approx60M_\odot$, and $A_{220}\approx 1$), we get $\zeta\approx 0.02$, and even for $\gamma=10$ the extra-mode amplitude ratio is small, $\hat A_{220}/ A_{220}\approx0.2$.

\section{BH ringdown amplitudescopy}
%
In this section, we present our results for the two case-study theories presented above.
The ringdown signal consists of two GW polarizations, with each component decomposed onto a basis of spin-weighted spheroidal harmonics that depend on the inclination angle $\iota$ of the remnant’s spin axis~\cite{Baibhav:2023clw}.

We focus on nonprecessing binaries and include one or two fundamental gravitational QNMs, namely $(220)$ and/or $(330)$, plus possibly the fundamental mode of the extra (scalar) field. Nevertheless, our methodology is general and can be extended to include additional higher-order modes, overtones, spin precession, and other field-induced modes~\cite{Crescimbeni:2024sam}. We exemplify our test on mock data by performing a Bayesian parameter estimation using the \texttt{PyCBC Inference} code infrastructure~\cite{Biwer:2018osg}.

\subsection{Ringdown Amplitudescopy with O4-like mock data and $\ell=m=2$ modes only}
\label{sec: ampl_O4}

Here, we consider a scenario where a ringdown signal in the modified gravity theories reported above is detected by current-generation interferometers with a representative ringdown optimal signal-to-noise ratio of ${\rm SNR}=20$. We assume detection by the LVK network operating at O4 sensitivity provided in Ref.~\cite{LIGO_T2200043_v3}, but the specific choice of the sensitivity curve has a mild impact on the analysis. We consider a GW250114-like system~\cite{LIGOScientific:2025epi}, with final mass $M=60M_{\odot}$ and final spin $\chi=0.67$. As a reference, the choice of SNR (computed starting at the peak of the ringdown signal) and parameters corresponds to a luminosity distance $d_L\approx 236\,{\rm Mpc}$ in O4.

For each of the two theories under consideration, we consider the following waveform models\footnote{Note that, for sources at low redshift --~including those detected to date and the loudest events expected in the future~-- the luminosity distance is degenerate with the overall ringdown amplitude. As a result, it can be omitted without loss of generality.}, all based on Eq.~\eqref{eq:rdmodelFINAL}:

\begin{itemize}
    \item {\texttt{SpecPA}: }\textit{Ordinary BH spectroscopy with the polar and axial modes.} According to the analysis of~\cite{Chung:2025wbg}, we include the fundamental polar $(220{\rm P})$ and axial $(220{\rm A})$ modes, using the QNMs provided in Refs.~\cite{Chung:2024vaf, Chung:2025gyg} as functions of $(M,\chi)$. This is possibly the easiest way to analyze modified gravity theories in the ringdown, since it involves only $\ell=m=2$ modes with no ordinary, highly-damped overtones.  The parameters that describe this model are
\begin{equation}
    \underline\theta=\{M,\chi,\ell_{\rm CS/GB},A_{220,P},A^R_{220,A},\phi_{220,P},\phi_{220,A},\iota\}   \,.\label{eq:param1}
\end{equation}
where $A^R_{220,A}=\frac{A_{220,A}}{A_{220,P}}$. This ratio is expected to be (significantly) smaller than unity, since the amplitude of the polar mode is expected to be (significantly) higher than the axial one in a merger, given that the latter are at least suppressed by some factor of the relevant source velocity in a post-Newtonian sense~\cite{Blanchet:2013haa}, as it happens for magnetic corrections relative to electric ones (see, e.g.,~\cite{Banihashemi:2018xfb,Abdelsalhin:2018reg} in the inspiral regime). However, lacking dedicated estimates for this ratio coming from actual simulations, we will keep it as a generic parameter. 

\item{\texttt{AmplPA}: } \textit{BH ringdown amplitudescopy for quadratic gravity theories.} This is the generalization of the \texttt{SpecPA} model presented above, but with the addition of the extra scalar mode. In this case, the waveform parameters are those in Eq.~\eqref{eq:param1} plus two more:
\begin{equation}
    \underline\theta \subset \{\gamma,\hat{\phi}_{220}^{s=0}\}
    \,.\label{eq:param1b}
\end{equation}
\end{itemize}

Since the modes used in this test have all the same spheroidal-harmonic decomposition, $\ell=m=2$, the inclination $\iota$ is degenerate with the total amplitude, so the above waveform models have effectively one parameter less.

For the results of this section, we assume nearly-symmetric binaries, so that the amplitude of putative odd higher modes can be neglected~\cite{Forteza:2022tgq}. Furthermore, we neglect the possible effect of $\ell=m=4$ modes. Although some evidence of the latter was identified in GW250114, they were poorly measured~\cite{LIGOScientific:2025obp}. Also, due to the lack of QNM computations and challenges in their modeling~\cite{Baibhav:2023clw,Berti:2025hly}, we neglect the addition of overtones, although a robust detection of $n=1$ was recently obtained for GW250114~\cite{LIGOScientific:2025obp}. 

In this first part, we perform an inference test by injecting a signal with \texttt{AmplPA} and recovering it with the same model. We consider the network configurations discussed above and inject a coupling length $\ell_{\rm CS/GB}=35\,{\rm km}$ which, for $M=60M_\odot$, corresponds to $\zeta\approx 0.024$. Although this value exceeds current upper bounds~\cite{Sanger:2024axs}, it represents a typical scale that can be constrained through BH spectroscopy in representative LVK events~\cite{Chung:2025wbg}. We also inject  $A^R_{220,A}=0.25$, and 
do not sample over the $\gamma$ parameter, fixing its value to the injected ones, $\gamma=\gamma_{\rm inj}\in\{0,1,5,10\}$. We chose to fix $\gamma$ for two reasons. The first is because, ideally, for a given theory, one could have an estimate of that value in terms of the progenitor parameters from simulations, which are finally available for certain theories.  The second is more technical: indeed, $\gamma$ is degenerate with final mass and $\alpha$, and its measurement might be difficult. 
Nonetheless, in Appendix~\ref{sec:corner_plots_ET} we relax this assumption and include $\gamma$ in the sampling parameters, obtaining similar results as those presented here.

\begin{figure*}[!t]
    \centering
    \includegraphics[width=0.46\linewidth]{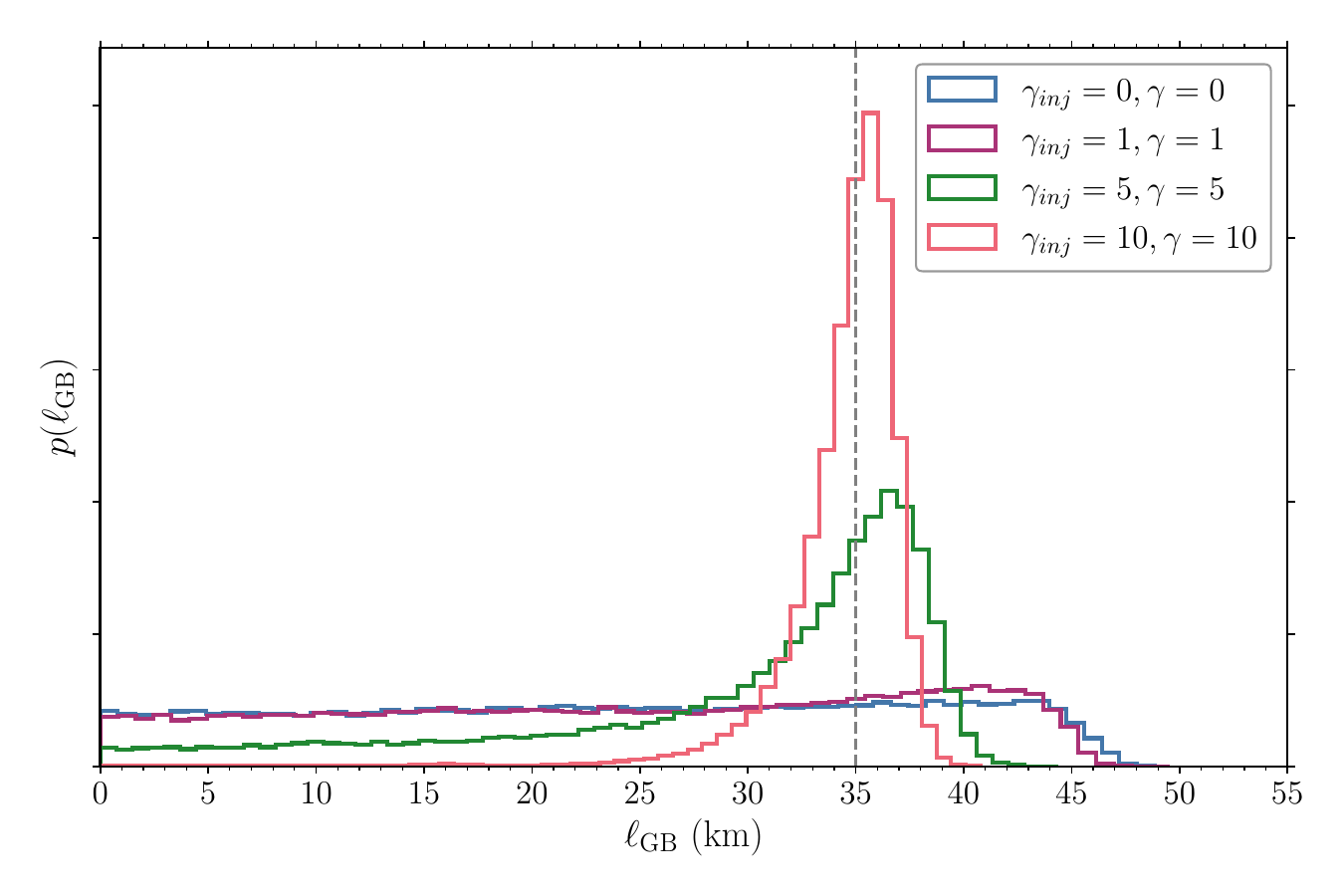}    \includegraphics[width=0.46\linewidth]{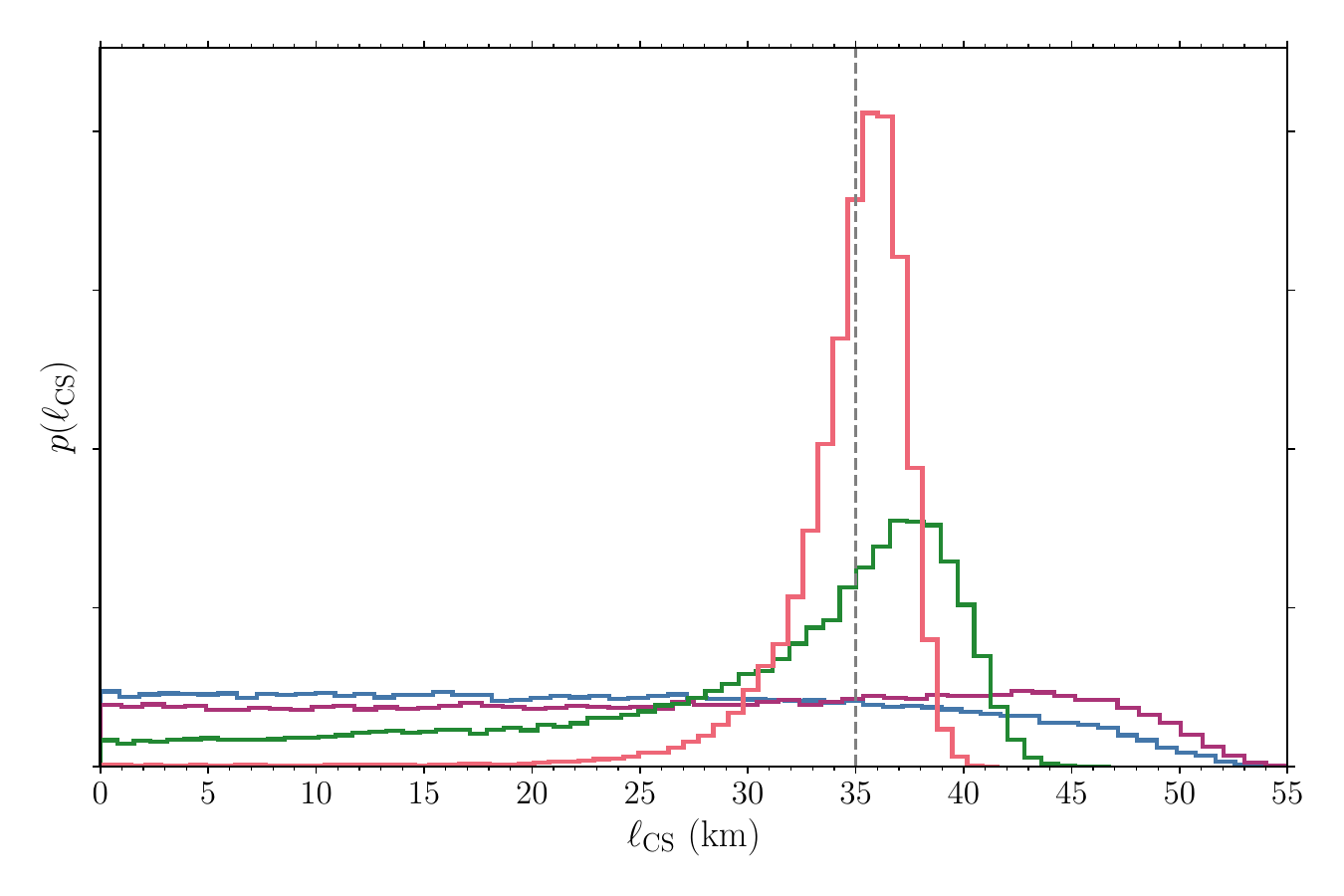}
    \includegraphics[width=0.46\linewidth]{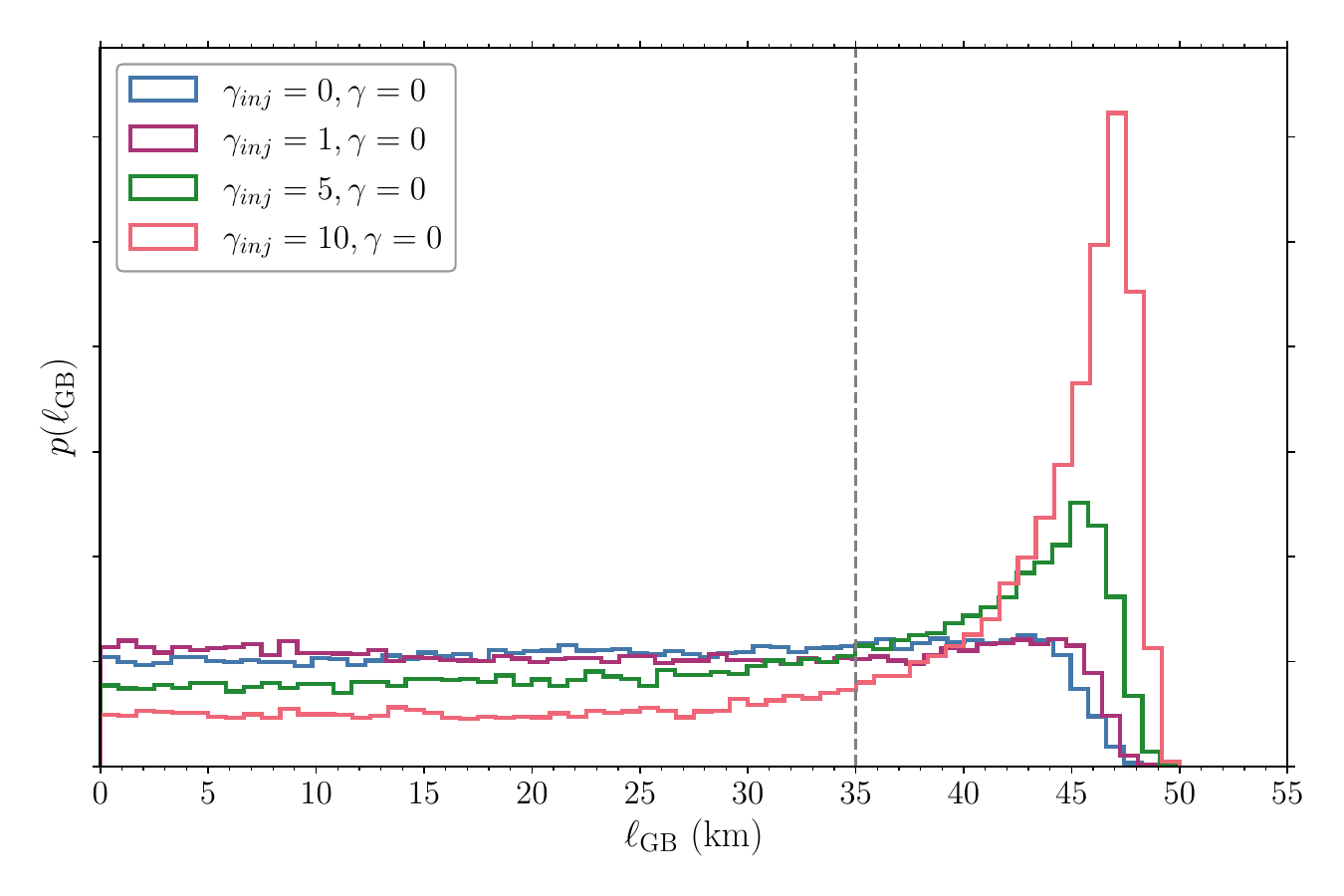}\includegraphics[width=0.46\linewidth]{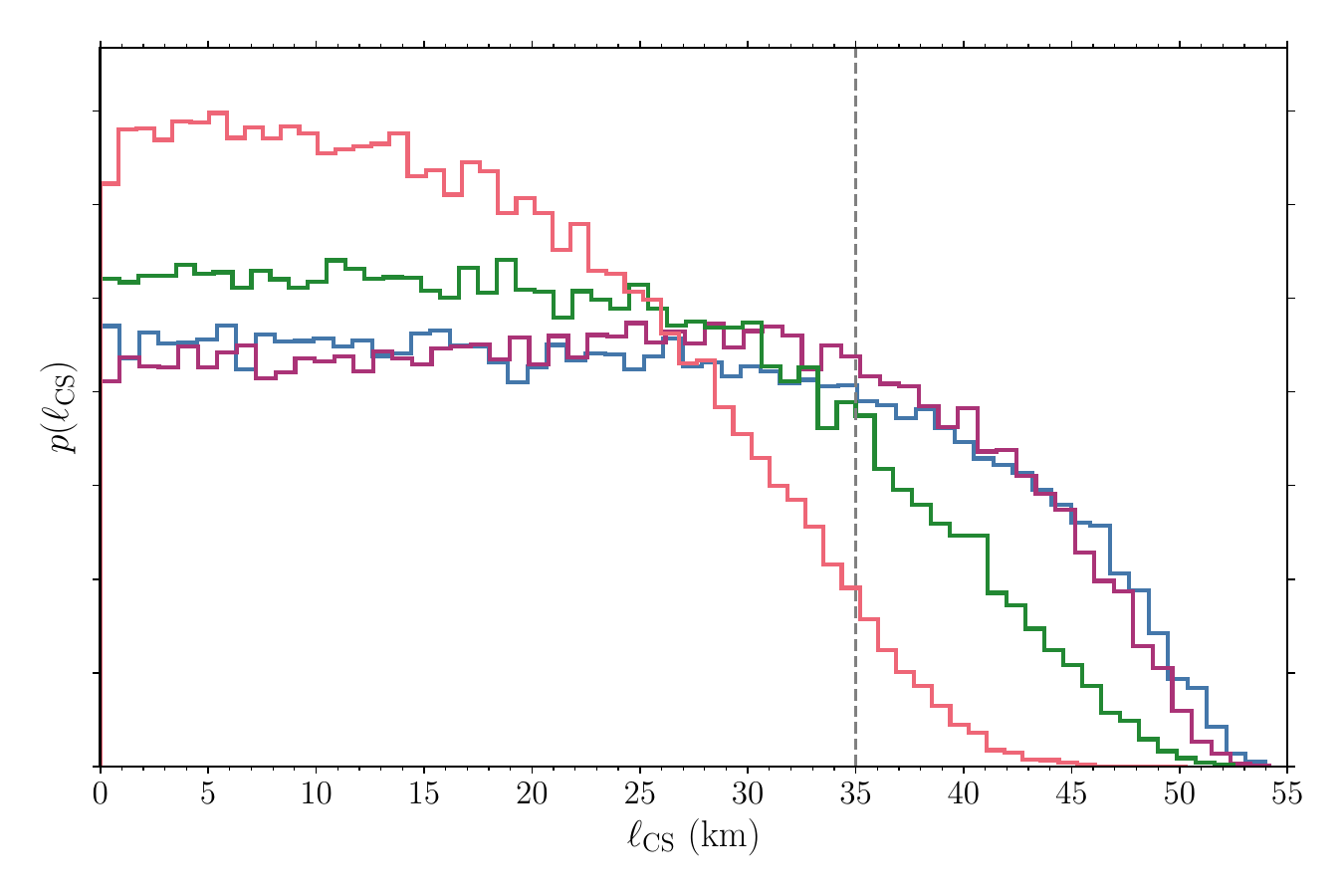}
    \caption{Distributions of $\ell_{\rm GB}$ (left panels) and $\ell_{\rm CS}$ (right panels) for $\frac{A_{220,A}}{A_{220_P}}=0.25$. Top and down panels represent the cases of $\gamma=\gamma_{\rm inj}$, and $\gamma=0$, respectively. We assume ${\rm SNR}=20$ with the O4 LVK network.  The prior on $\ell_{\rm GB/CS}$ is uniform in the range $[0,60]\,{\rm km}$.} \label{fig:alpha_0_3_cases}
\end{figure*}

The resulting posteriors for $\ell_{\rm CS}$ and $\ell_{\rm GB}$ are shown in the top panels of Fig.~\ref{fig:alpha_0_3_cases}. As $\gamma$ increases, the constraints on the coupling constant become tighter. The cases with $\gamma=0$ and $\gamma=1$ give nearly identical results, reflecting the fact that the scalar amplitude is too small to be relevant in that regime. Even though our analysis is based on mock data, we can compare the $\gamma=0$ case with the results of Ref.~\cite{Chung:2025wbg} for GW150914. In that work, the reported upper bounds on $\ell_{\rm CS/GB}$ are 53.6\,km and 46.5\,km, while our analysis yields 45.6\,km and 43.8\,km for the two theories. The overall agreement is satisfactory, while the remaining differences can be ascribed to the use of mock data with a {\it total post-merger} ${\rm SNR}=20$ in this work, as opposed to real data with ringdown ${\rm SNR}\approx 8$ in Ref.~\cite{Chung:2025wbg}.

Next, we assess the impact of neglecting a scalar mode in the inference.  
To this end, we inject signals with \texttt{AmplPA} for different values of $\gamma$, and recover them with \texttt{SpecPA}. In this setup, the true signal includes an additional scalar mode, while the recovery template accounts only for the polar and axial gravitational modes. To quantify the mismatch, we compute the Bayes factors
\begin{equation}
\mathcal{B}(\gamma) =
\frac{Z_\texttt{SpecPA}}{Z_\texttt{AmplPA}} \,,
\label{BFs_SNR=20}
\end{equation}
where the numerator denotes the evidence of a recovery with $\gamma=0$, whereas the denominator denotes the evidence of a recovery with the true injected value.  
The posterior distributions for $\ell_{\rm CS/GB}$ in both cases are shown in Fig.~\ref{fig:alpha_0_3_cases}, while the corresponding Bayes factors are listed in Table~\ref{tab:BF_table_SNR_20}. As expected, the Bayes factors increase monotonically with $\gamma$. According to Jeffreys’ scale~\cite{Jeffreys:1939xee}, a $\log_{10}$ Bayes factor greater than $-1$ ($-2$) indicates strong (decisive) evidence in favor of the true model. In our case, we find strong evidence in both theories for $\gamma=10$, while for $\gamma\lesssim5$ the extra scalar mode has negligible statistical evidence.

Finally, from the bottom panels of Fig.~\ref{fig:alpha_0_3_cases} we observe that neglecting the scalar mode in the recovery biases the inferred coupling length: $\ell_{\rm CS}$ is underestimated, while $\ell_{\rm GB}$ is overestimated, with respect to the injected value of $35\,{\rm km}$. The opposite shifts in the distributions when $\gamma_{\rm inj}$ is large can be understood as follows. The $(220)$ scalar mode has higher frequency than the $(220)$ axial and polar modes in both GB or CS theories~\cite{Crescimbeni:2024sam}. From Fig.~\ref{fig:dCS_deviations}, we see that for $\chi\approx 0.67$ the GB deviation is positive, whereas the CS is negative. Hence, in the former case one would need a larger coupling to recover the scalar mode with a $\gamma=0$ model, while in the latter case one would need $\ell_{\rm CS}\propto\alpha_{\rm CS}<0$, which is clearly forbidden. Hence, the smallest allowed value, $\ell_{\rm CS}=0$, is favored in order to make the (negative) CS deviation of the $(220)$ axial mode as small as possible. Both trends are consistent with the posteriors shown in the bottom panels of Fig.~\ref{fig:alpha_0_3_cases}.

\begin{table}[h!]
\centering
\begin{tabular}{|c|c|c|c|}
\hline
Theory & $\log_{10}\mathcal{B}(\gamma=1)$ & $\log_{10}\mathcal{B}(\gamma=5)$ & $\log_{10}\mathcal{B}(\gamma=10)$ \\
\hline
CS & -0.04 & -0.52  & -1.77 \\
GB & -0.03 & -0.28  & -1.70 \\
\hline
\end{tabular}
\caption{log$_{10}$ Bayes factors defined in Eq.~\eqref{BFs_SNR=20} for CS (first row) and GB (second row) theories for $\frac{A_{220,A}}{A_{220_P}}=0.25$.
}
\label{tab:BF_table_SNR_20}
\end{table}

We also have performed the analyses of Fig.~2 by sampling with a uniform prior on $\gamma$, and we report the results in Appendix~\ref{sec:corner_plots_ET} for the case of CS, showing that sampling on $\gamma$ has a small impact on the results.

These results demonstrate that, even at moderate SNRs accessible to the LVK network in its current configuration, neglecting scalar modes can introduce modeling errors that bias the inferred coupling constant. Such biases may in turn mimic spurious deviations from GR or conceal genuine ones~\cite{Gupta:2024gun}.

\subsection{Ringdown Amplitudescopy with higher modes and future ringdown tests}
\label{sec: ampl_ET}

Here we perform a forecast analysis assuming an event with ringdown ${\rm SNR}=80$ measured by a third-generation interferometer such as the Einstein Telescope~(ET)~\cite{Hild:2010id, Maggiore:2019uih, Branchesi:2023mws, Abac:2025saz}.
We optimistically consider a light remnant with $M=20M_{\odot}$, which provides stronger constraints on dimensionful couplings as those arising in quadratic gravity~\cite{Maselli:2019mjd}. As a reference, this corresponds to a luminosity distance $d_L\approx 437\,{\rm Mpc}$ in ET.
We again consider the final spin $\chi=0.67$ as a representative value. We assume ET in a triangle configuration with arm length of 15km, with the PSD provided in Ref.~\cite{Branchesi:2023mws}. Also in this case, the impact of the different PSDs has a mild impact on the analysis, and the impact of the BH spectroscopy results due to the choice of the detector configuration is discussed in Ref.~\cite{Branchesi:2023mws}.

Instead of considering axial and polar $\ell=m=2$ modes as done in the previous section, we perform a more standard ringdown analysis with the (2,2,0), (3,3,0) modes, possibly supplemented by the extra scalar mode.
We consider polar modes only, since the amplitude of axial modes is unknown and source dependent; therefore, at least for certain systems, the fundamental polar $\ell=m=3$ is expected to be the second most constrained gravitational mode in the ringdown signal~\cite{Bhagwat:2023jwv}.

Our ringdown model is again based on Eq.~\eqref{eq:rdmodelFINAL} with the following parameters:
    \begin{equation}
\underline\theta=\{M,\chi,\alpha,A_{220},A_{R,330},\phi_{220},\phi_{330},\gamma,\hat\phi_{220},\iota\}   \,,\label{eq:param3}
\end{equation}
where we denote the amplitude ratio $A_{R,330}=\frac{A_{330}}{A_{220}}$. Note that in this case the inclination $\iota$ is relevant, as it affects the effective relative amplitude of the modes in the signal. In these injections, we consider the representative value $\iota\approx\frac{\pi}{8}$, which respects the selection effects in observing mostly nearly face-on signals~\cite{Usman:2018imj}.

\begin{figure*}[!t]
    \centering
    \includegraphics[width=0.9\linewidth]{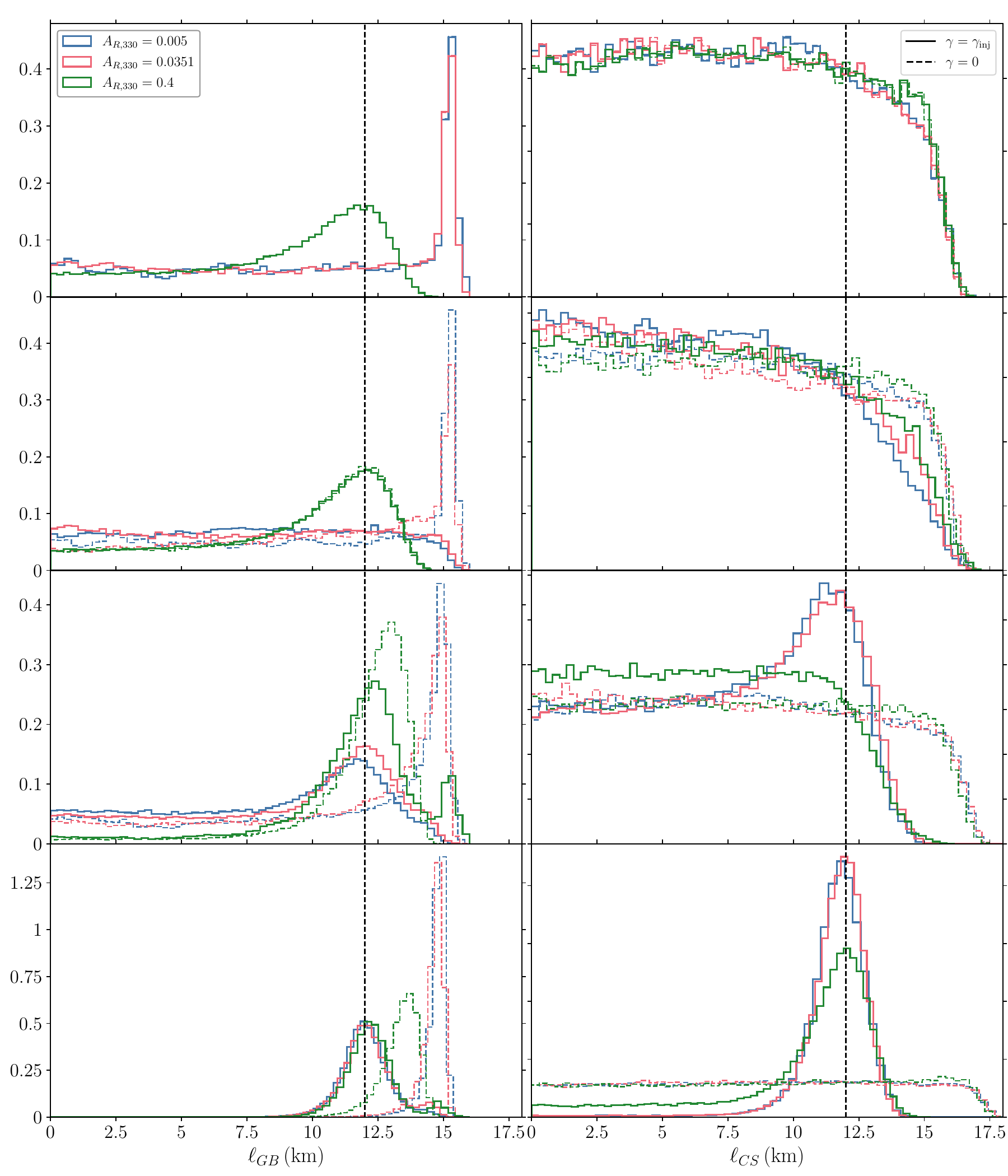}
    \
    \caption{Distributions of $\ell_{\rm GB}$ (left panels) and $\ell_{\rm CS}$ (right panels) for $A_{R, 330}=\frac{A_{330}}{A_{220}}=[0.005,0.351,0.4]$ (in blue, red, and green, respectively). Each row refers to a different value of injected $\gamma$ (from top to bottom: $\gamma_{\rm inj}=0,1, 5, 10$. The continuous and dashed distributions correspond to a recovery with $\gamma=\gamma_{\rm inj}$ and $\gamma=0$, respectively. In the GB case, the priors are uniform across $\ell_{\rm GB}\in[0,16]\,{\rm km}$ and $M\in[15,40]M_{\odot}$ while in the CS case they are uniform across $\ell_{\rm CS}\in[0,20]\,{\rm km}$ and $M\in[17,40]M_{\odot}$.  
}
    \label{fig:330}
\end{figure*}

The above model with the parameters in ~\eqref{eq:param3} includes the case of ordinary BH spectroscopy (when $\gamma=0$) and the case in which only a single ordinary mode dominates the inference ($A_{R,330}\rightarrow0$). In particular, this last condition can be confidently achieved when the progenitors have nearly-equal masses, thus suppressing the higher modes~\cite{Forteza:2022tgq}. Interestingly, even in the latter case a ringdown test is still possible if $\gamma\neq0$, since the presence of an extra mode breaks the degeneracy between $(\delta f_{220},\delta\tau_{220})$ and $(M,\chi)$, without the need of including either higher-multipole modes or overtones~\cite{Crescimbeni:2024sam}.

In Fig.~\ref{fig:330}, we show the posterior distributions of the coupling constant $\ell_{\rm GB}$ (left panels) and $\ell_{\rm CS}$ (right panels) for different choices of $A_{R,330}$ \footnote{Notice that the values of $A_{\rm R,330}$ reach values lower than 0.04, corresponding to mass ratio very close to unity. Even if these values cannot be considered as a confident detection of the 330 mode, we report it anyway, as we can analyze the impact of adding a third mode with different excitations.} and $\gamma_{\rm inj}$, injecting $\ell_{\rm GB/CS}=12\,{\rm km}$, while the full corner plots are reported in Appendix~\ref{sec:corner_plots_ET}. Compared to the previous analysis, we can consider a smaller coupling, as the remnant mass is smaller in this case, corresponding to $\zeta\approx 0.027$.  
In each panel, we show the results of two analyses: 1)~recovery with a model where $\gamma=\gamma_{\rm inj}$ (continuous histograms), and 2)~recovery with a model where $\gamma=0$ (dashed histograms), i.e. analog to \texttt{SpecPA} of the previous section but using the $(220)$ and $(330)$ polar modes rather than the $(220)$ axial and polar modes.

The first row of Fig.~\ref{fig:330} corresponds to a standard ringdown analysis ($\gamma_{\rm inj}=0$). As expected, when the secondary mode is weakly excited (small $A_{R,330}$), the posterior is almost flat and tends to coincide with the uniform prior $\ell_{\rm GB}\in[0,16]\,{\rm km}$ and $\ell_{\rm CS}\in[0,20]\,{\rm km}$, respectively \footnote{We did not consider to higher values of $\ell_{\rm GB}$ due to stability issue of the simulations related to the frequency shifts reported in Ref.~\cite{Chung:2024vaf}.} . However, we find some interesting behaviors for both the theories. For GB, we find that the distribution in $\ell_{\rm GB}$ shifts near the upper bound of the prior for low $A_{330,R}$ values. We suspect that this behavior is given by a secondary peak in the final mass distribution (see Appendix~\ref{sec:corner_plots_ET}), which might be due to a poor convergence of the analysis when the 330 mode has a low amplitude. Indeed, since the beyond-GR effects depend on the combination $\zeta=\alpha^2/M^4$, a bias in the mass correlates with a bias in the coupling.
However, this bias is not an actual problem, given the low value of the injected $A_{R,330}$ and the fact that the recovery is correct for more realistic values of $A_{R,330}$. For CS instead, an upper bound at $\ell_{\rm CS}\approx18$km is present, because the posterior distribution of $A_{R,330}$ peaks at values higher than the injection, leading to a non-negligible effect of the coupling.
When the secondary-mode amplitude is large ($A_{R,330}=0.4$), the injected value of the coupling is well recovered in GB theory, whereas this is not the case for CS theory, since the CS corrections to the $(330)$ polar mode are smaller (see Fig.~\ref{fig:dCS_deviations}). As a reference, for non-precessing binaries, $A_{R_,330}\lesssim 0.5$ for all the mass-ratio and spin configurations~\cite{Forteza:2022tgq}.

The second row of Fig.~\ref{fig:330} shows that the situation with $\gamma_{\rm inj}=1$ is very similar to the $\gamma_{\rm inj}=0$ case, again because the extra scalar mode is not sufficiently excited to impact the posterior of the coupling. Nevertheless, we see that the unbiased distribution corresponding to $\gamma=\gamma_{\rm inj}$ and the biased one corresponding to $\gamma=0$ start differing from each other, because in the latter case the recovery is done with the wrong model.

The role of the scalar mode becomes more prominent for $\gamma_{\rm inj}\gtrsim5$. In this case, we see various effects: i)~the difference between the biased and unbiased distributions is more evident; ii) only the unbiased model can correctly recover the injected value, and, for GB theory, the posterior of the biased model peaks at the wrong value, as expected; iii)~when the extra scalar mode becomes important, the posteriors are informative also in the CS case.
These trends are more evident in the last row of Fig.~\ref{fig:330} ($\gamma_{\rm inj}=10$). In this case, only the unbiased model recovers the correct value and, remarkably, this happens also when $A_{R,330}$ is small. As previously mentioned, the $(330)$ mode is not needed in this case to constrain the theory, since the extra scalar mode can be instead used, together with the fundamental gravitational $\ell=m=2$ mode, for a standard spectroscopy test~\cite{Crescimbeni:2024sam}.
When the scalar mode dominates (namely, when either $A_{R,330}$ is small or $\gamma$ is large), the constraints on CS and GB theories are similar, since this mode is minimally coupled and hence model independent.

Indeed, we expect that, for any $\gamma \neq 0$, there exists a critical threshold of $A_{R,330}$ below which the extra scalar mode inevitably dominates over the $\ell = m = 3$ mode. Overall, as clear from 
Fig.~\ref{fig:330}, when $\gamma\gtrsim5$ the bounds on the coupling are significantly better than when neglecting the extra mode, in agreement with the findings of~\cite{Crescimbeni:2024sam}.

This shows that focusing only on QNM frequency shifts, while ignoring the scalar-mode amplitudes, significantly weakens the ability to constrain the extra degrees of freedom predicted by generic beyond-GR theories~\cite{Lestingi:2025jyb}. 

Finally, we also compute the Bayes factors between the biased model and the unbiased model in this case. While the posteriors of the unbiased model shown in Fig.~\ref{fig:330} are clearly more accurate for both the GB and the CS cases, we observe that this does not always result in a more discriminating Bayes factor.
In the CS case, the Bayes factors grow (in absolute value) monotonically either as $\gamma$ increases or $A_{R,330}$  decreases, as expected. However, in the GB case, the likelihood of the biased distribution is as high as that of the unbiased one, resulting in $\log_{10}{\cal B}\approx 0$ even when $\gamma=10$ and $A_{R,330}=0.005$. This suggests that it would be hard to exclude statistically the biased model, at least based on the Bayes factor estimators. In this case, the maximum likelihood parameters are biased with respect to the injected values, but they still perform a good fit on the model, thus returning an unfaithful Bayes factor, as also observed in other situations~\cite{Isi:2021iql}. This feature for GB theory might be due to the fact that the posterior of $\ell_{GB}$ is constrained near the prior upper bound, which is close to the injected value. As a future extension, it might be interesting to understand if the small values of $\log_{10}(\mathcal{B})$ persist when adding \textit{further} modes in addition to 330P. For instance, it would be interesting to merge the analyses performed in Sections~\ref{sec: ampl_O4} and~\ref{sec: ampl_ET}, thus including 220P+220A+330P plus, possibly, a scalar mode. More in general, an accurate forecast should be performed, to understand which are the modes that impact the most in the analysis and which ones can mask the scalar contribution.

\section{Conclusions}

Our analysis highlights the relevance of BH ringdown amplitudescopy, namely including extra field-induced modes in ringdown templates. The presence of these modes is unavoidable in beyond-GR theories but has been neglected in the majority of current searches.

We have considered two case studies of gravity theories with quadratic curvature corrections (CS and GB theories), finding that a sufficiently excited scalar mode would affect the constraints on the coupling constant of the theory. On the one hand, if properly included in the ringdown analysis, an extra mode makes the upper bounds on GR deviations more stringent. However, if not included in the analysis, the absence of an extra mode can introduce significant biases in the measurements due to mismodelling of beyond-GR effects~\cite{Gupta:2024gun}.

In this work, since we have injected a ringdown waveform, we are neglecting the issue of the amplitude stability~\cite{Cheung:2023vki}, as the true signal does not contain nonlinearities and transient effects. However, when performing real data analysis, the amplitude stability is reached only at late times. In this last case, there can be two ways of performing the \textit{amplitudescopy} analysis: either starting at a fixed time sufficiently far after the merger (see, for instance, in~\cite{Chung:2025wbg}), or improving currently beyond-GR IMR waveform models (e.g.~\cite{Julie:2024fwy}). We leave these improvements for future works.

Further, as a proof of principle, here we focused on extra scalar modes and non-precessing binaries, but considering other fields (e.g., vectors, tensors), precessing binaries, or modes other than $\ell=m=2,3$ are straightforward extensions~\cite{Crescimbeni:2024sam}.
It would also be interesting to consider the effect of extra \emph{massive} degrees of freedom. Even if the latter do not propagate at relatively low frequencies, they can nevertheless excite the gravitational response and leave an imprint in the ringdown (see~\cite{Cardoso:2020nst,Antoniou:2024jku} for some recent examples).

Finally, the most important follow-up work is to quantify the excitation amplitudes of extra modes in the ringdown for a given theory, which can be achieved using recent numerical-relativity simulations beyond GR (e.g.,~\cite{Okounkova:2019zjf,Okounkova:2020rqw,East:2020hgw,East:2021bqk,Figueras:2021abd,AresteSalo:2022hua,Corman:2022xqg,Cayuso:2023aht, Doneva_2025_prep}) or point-particle models~\cite{Blazquez-Salcedo:2016enn, Silva:2024ffz}.


{\bf Software.}
Inference simulations have been carried out with \texttt{pycbc inference}~\cite{Biwer:2018osg}.
The manuscript content has been derived using publicly available software: {\tt matplotlib}, {\tt corner}, {\tt json}, {\tt numpy}~\cite{Hunter:2007, corner, bray2014javascript, harris2020array}. Codes are available upon request.

\begin{acknowledgements}
We thank Juan Calderon Bustillo, Gregorio Carullo, Giovanni D'Addario, Matteo Della Rocca, Cecilia Maria Fabbri, Stephen Green, Jacopo Lestingi, Laura Sberna, and Thomas Sotiriou for interesting discussions. P.P. and F.C. are partially supported by the MUR FIS2 Advanced Grant ET-NOW (CUP:~B53C25001080001) and by the INFN TEONGRAV initiative. F.C. acknowledges the financial support provided under the “Progetti per Avvio alla Ricerca Tipo 1”, protocol number AR12419073C0A82B.
X.J. is supported by the Spanish Ministerio de Ciencia, Innovación y Universidades (Beatriz Galindo, BG22-00034)
and cofinanced by UIB; the Spanish Agencia Estatal de Investigación
Grants No. PID2022-138626NB-I00, No. RED2022-
134204-E, and No. RED2022-134411-T, funded by
MCIN/AEI/10.13039/501100011033/FEDER, UE; the
MCIN with funding from the European Union NextGenerationEU/PRTR (No. PRTR-C17.I1); the Comunitat
Autonòma de les Illes Balears through the Direcció General de Recerca, Innovació I Transformació Digital with
funds from the Tourist Stay Tax Law (No. PDR2020/11
- ITS2017-006), and the Conselleria d’Economia, Hisenda
i Innovació Grant No. SINCO2022/6719.
Some numerical computations have been performed at the Vera cluster supported by the Italian Ministry of Research and by Sapienza University of Rome, with the University of Birmingham's BlueBEAR and the computer resources at MareNostrum5 (RES-AECT-2025-2-0038).
\end{acknowledgements}

\appendix

\section{Frequency shift computations}
\label{sec:fs_computation}
In this appendix, we provide the computation of the frequency shifts described in Refs.~\cite{Chung:2024vaf, Chung:2025gyg}. We write the beyond-GR correction to the complex frequency in geometric units as
\begin{equation}
\omega_i(\chi)=\omega_{i,0}(\chi)+\zeta\,\omega_{i,1}(\chi),
\end{equation}
where $\omega_{i,0}$ is the GR QNM complex frequency, and $\omega_{i,1}$ is the first-order correction, which we decompose into real and imaginary parts,
\begin{equation}
\omega_{i,0}=\Re(\omega_{i,0})+i\,\Im(\omega_{i,0}),\qquad
\omega_{i,1}=\Re(\omega_{i,1})+i\,\Im(\omega_{i,1}).
\end{equation}
We define the spin-dependent functions of Eq.~\eqref{corrections} as
\begin{equation}
F_i(\chi)\equiv \frac{\Re(\omega_{i,1}(\chi))}{\Re(\omega_{i,0}(\chi))},
\qquad
T_i(\chi)\equiv \frac{\Im(\omega_{i,1}(\chi))}{\Im(\omega_{i,0}(\chi))},
\end{equation}
so that
\begin{equation}
\Re(\omega_{i,1})=F_i(\chi)\,\Re(\omega_{i,0}),\qquad
\Im(\omega_{i,1})=T_i(\chi)\,\Im(\omega_{i,0}).
\end{equation}
It follows that the corrected real and imaginary parts can be written as
\begin{align}
\Re(\omega_i) &= \Re(\omega_{i,0})\bigl[1+\zeta\,F_i(\chi)\bigr],\\
\Im(\omega_i) &= \Im(\omega_{i,0})\bigl[1+\zeta\,T_i(\chi)\bigr].
\end{align}
The physical frequency and damping time are defined by
\begin{equation}
f_i \equiv \frac{\Re(\omega_i)}{2\pi},\qquad
\tau_i \equiv -\frac{1}{\Im(\omega_i)},
\end{equation}
with the GR values
\begin{equation}
f_{i,0}=\frac{\Re(\omega_{i,0})}{2\pi},\qquad
\tau_{i,0}=-\frac{1}{\Im(\omega_{i,0})}.
\end{equation}
Therefore,
\begin{align}
f_i &= f_{i,0}\bigl[1+\zeta\,F_i(\chi)\bigr],\\
\tau_i &= \tau_{i,0}\,\frac{1}{1+\zeta\,T_i(\chi)}.
\end{align}
Defining the relative deviations
\begin{equation}
\delta f_i \equiv \frac{f_i}{f_{i,0}}-1,\qquad
\delta\tau_i \equiv \frac{\tau_i}{\tau_{i,0}}-1,
\end{equation}
we obtain the exact relations
\begin{align}
\delta f_i &= \zeta\,F_i(\chi),\\
\delta\tau_i &= -\,\frac{\zeta\,T_i(\chi)}{1+\zeta\,T_i(\chi)}
\end{align}
which are the terms of Eq.~\eqref{corrections}. The first-order complex correction $\omega_{i,1}(\chi)$ is expressed as a polynomial in spin:
\begin{equation}
\omega_{i,1}(\chi) = \sum_{j=0}^{8} \omega^{(i)}_{j}\,\chi^{j},
\qquad
\omega^{(i)}_{j} = a^{(i)}_{j} + i\,b^{(i)}_{j},
\end{equation}
where the complex coefficients $\omega^{(i)}_{j}$ are obtained from the fits in Table~V of Ref.~\cite{Chung:2024vaf} for GB, and Table~V of Ref.~\cite{Chung:2025gyg} for CS.
The tables below report the full complex coefficients $\omega^{(i)}_{j}$ for the dominant $(2,2,0)$ and $(3,3,0)$ modes in CS and GB, separately for the axial and polar parities.

\begin{table*}[t]
\centering
\setlength{\tabcolsep}{6pt}
\renewcommand{\arraystretch}{1.15}
\scriptsize
\begin{tabular}{c c c c c}
\hline\hline
& \multicolumn{2}{c}{$220$} & \multicolumn{2}{c}{$330$} \\
\hline
$j$ & axial & polar & axial & polar \\
\hline
0 & $-0.246041 - 0.125482i\,i$ & $0$ & $-0.912752-0.164615\,i$ & $0$ \\
1 & $-0.35908-0.184144\,i$ & $-0.0307839+ 0.017411\,i$ & $-1.09512-0.0911042\,i$ & $-0.0901535-0.000837525\,i$ \\
2 & $-0.595672+0.0876598\,i$ & $0.013835-0.214353\,i$ & $-1.23272+0.116646\,i$ & $-0.0763526-0.149877\,i$ \\
3 & $1.41373+ 0.258192\,i$ & $0.398761- 0.447376$ & $2.37515+0.763688\,i$ & $0.452064-0.248466\,i$ \\
4 & $-11.2668 + 0.703524\,i$ & $-6.66087 + 0.448166\,i$ & $-16.7427-0.927771\,i$ & $-6.70664-0.508568\,i$ \\
5 & $42.008-1.28144\,i$ & $24.7362- 1.091\,i$ & $52.5812 + 1.12866\,i$ & $25.7745 + 1.18438\,i$ \\
6 & $-82.8112 + 1.05377\,i$ & $-55.9612 + 2.5070\,i$ & $-97.6497 + 0.865323\,i$ & $-60.5245-1.60982\,i$ \\
7 & $83.7508-0.833429\,i$ & $63.124-1.10815\,i$ & $97.2953-1.91376\,i$ & $-60.5245-1.60982\,i$ \\
8 & $-34.8784$ & $-28.7982$ & $-40.6093$ & $-32.9995$ \\
\hline\hline
\end{tabular}
\caption{Complex coefficients $\omega^{(i)}_j$ for the $(220)$ and $(330)$ modes and axial/polar parities in CS theory.}
\label{tab:dcs_omega_coeffs_cols}
\end{table*}

\begin{table*}[t]
\centering
\setlength{\tabcolsep}{6pt}
\renewcommand{\arraystretch}{1.15}
\scriptsize
\begin{tabular}{c c c c c}
\hline\hline
& \multicolumn{2}{c}{$220$} & \multicolumn{2}{c}{$330$} \\
\hline
$j$ & axial & polar & axial & polar \\
\hline
0 & $0.055241+0.00686713\,i$ & $-0.215202-0.0734094\,i$ & $0.109706+0.0061152\,i$ & $-0.872789-0.113506\,i$ \\
1 & $0.985294+0.0636233\,i$ & $-2.51816-0.411031\,i$ & $0.685561+0.231546\,i$ & $-1.20198-0.339974\,i$ \\
2 & $-18.4902+0.325569\,i$ & $48.4695+9.59898\,i$ & $-11.6733-4.20702\,i$ & $2.82484+9.09087\,i$ \\
3 & $157.802-4.46276\,i$ & $-431.428-82.4765\,i$ & $108.698+37.6761\,i$ & $-35.6024-75.7708\,i$ \\
4 & $-682.224+19.8274\,i$ & $1935.01+378.832\,i$ & $-510.61-178.723\,i$ & $165.435+341.913\,i$ \\
5 & $1660.08-55.4163\,i$ & $-4831.15-965.058\,i$ & $1334.38+467.415\,i$ & $-434.981-857.681\,i$ \\
6 & $-2307.78+87.744\,i$ & $6808.99+1386.33\,i$ & $-1970.88-687.899\,i$ & $638.737+1219.16\,i$ \\
7 & $1711.32-73.0696\,i$ & $-5065.1-1050.91\,i$ & $1536.55+532.083\,i$ & $-488.58-915.499\,i$ \\
8 & $-526.516+25.0324\,i$ & $1544.32+325.905\,i$ & $-493.701-167.701\,i$ & $150.043+280.837\,i$ \\
\hline\hline
\end{tabular}
\caption{Same of Table~\ref{tab:dcs_omega_coeffs_cols}, but for GB theory.}
\label{tab:esgb_omega_coeffs_cols}
\end{table*}

\section{Additional results}
\label{sec:corner_plots_ET}
\begin{figure}[!t]
    \centering
    \includegraphics[width=0.96\linewidth]{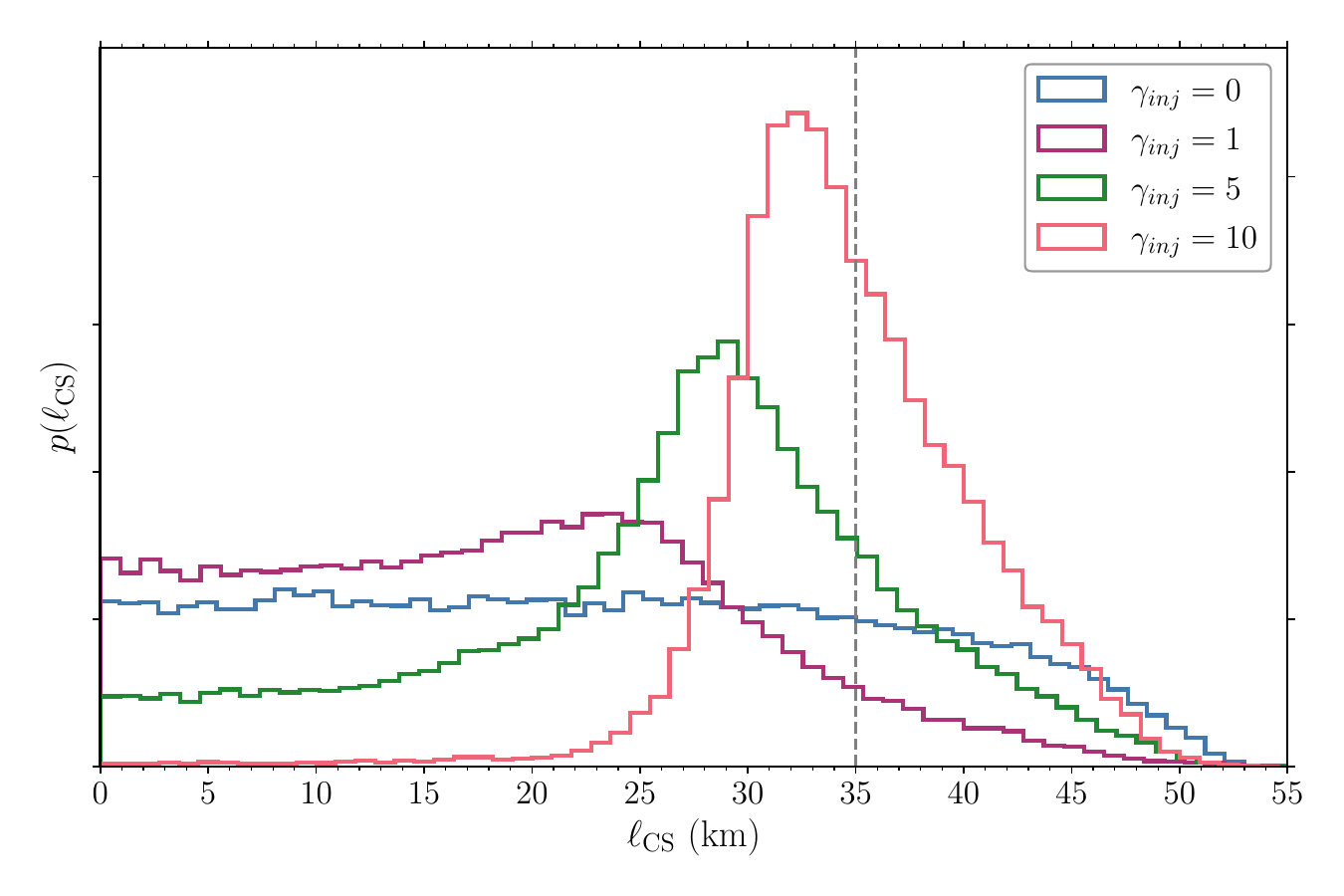}
    
    \caption{Same as top left panel of Fig. \ref{fig:alpha_0_3_cases}, but sampling on $\gamma$.} \label{fig:sampled_gamma_fig2}
\end{figure}

\begin{figure*}[p]
  \centering
  \includegraphics[width=0.9\textwidth]{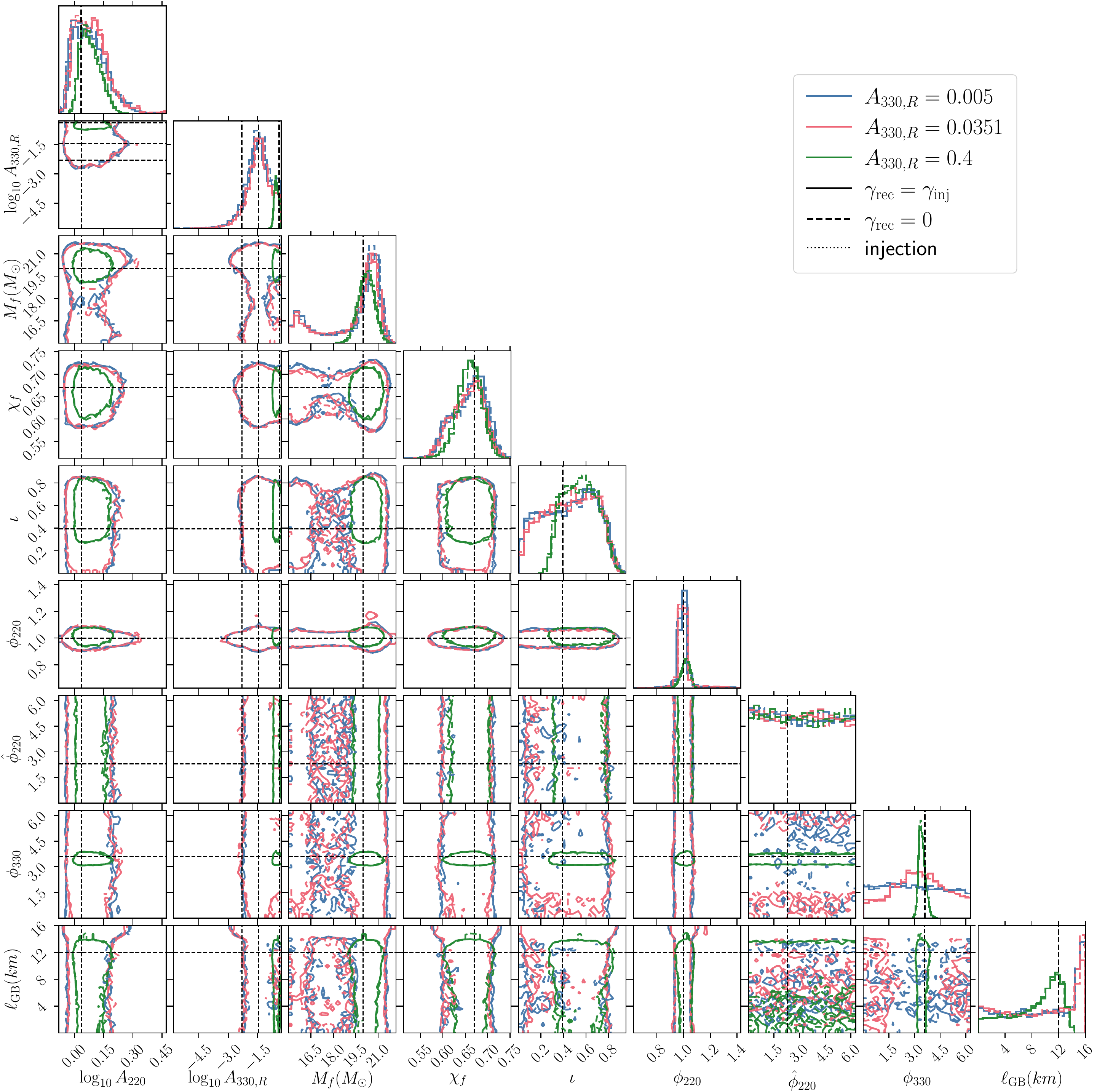}
  \caption{Corner plot for GB 220P+330P injection at ${\rm SNR}=80$ with $\sqrt{\alpha}=12$ and $\gamma_{\rm inj}=0$.
  Curves show posteriors for three injected amplitudes $A_{330}\in\{0.005,\,0.0351,\,0.4\}$ (blue/red/green).
  Solid curves correspond to recovery with $\gamma=\gamma_{\rm inj}$, dashed curves to recovery with $\gamma=0$.
  Black dashed lines indicate injected values. For the 2D distributions, we report the 90\% credible level.}
  \label{fig:corner-esgb-phi0}
\end{figure*}

\begin{figure*}[p]
  \centering
  \includegraphics[width=0.9\textwidth]{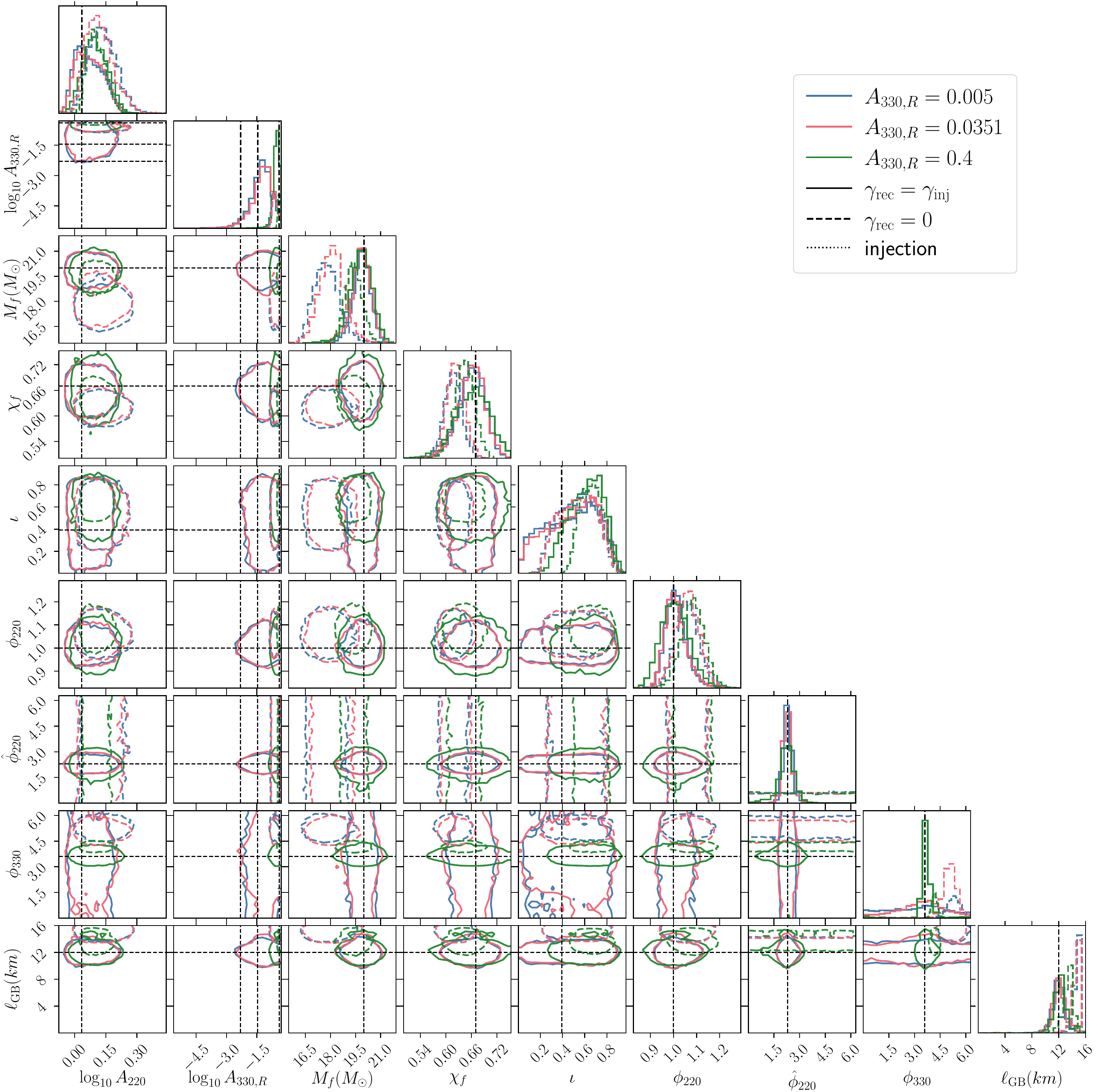}
  \caption{Same as Fig. \ref{fig:corner-esgb-phi0}, but for $\gamma_{\rm inj}=10$.}
  \label{fig:corner-esgb-phi10}
\end{figure*}

\begin{figure*}[p]
  \centering
  \includegraphics[width=0.9\textwidth]{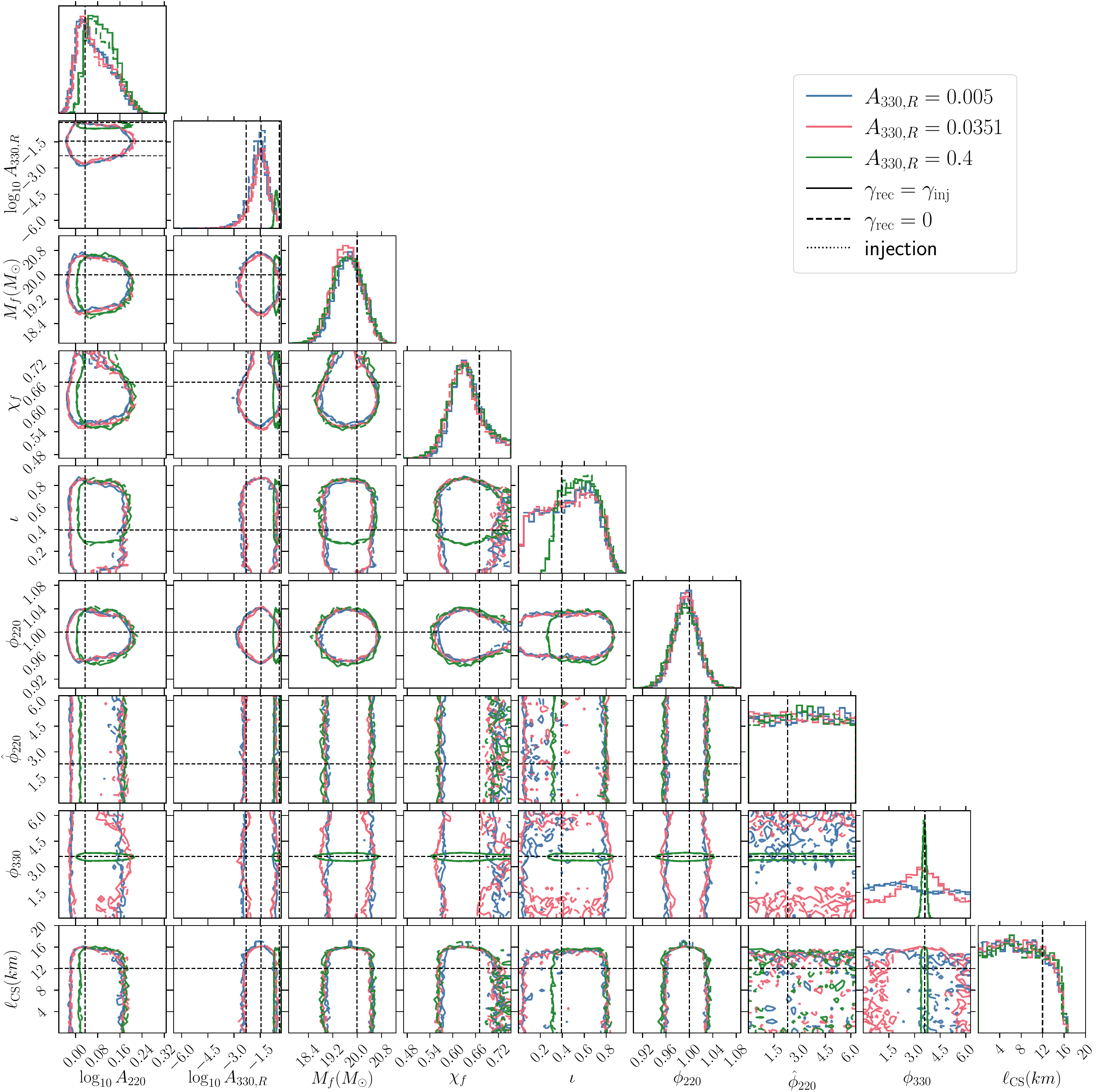}
  \caption{Same as Fig. \ref{fig:corner-esgb-phi0}, but for CS and $\gamma_{\rm inj}=0$.}
  \label{fig:corner-dcs-phi0}
\end{figure*}

\begin{figure*}[p]
  \centering
  \includegraphics[width=0.9\textwidth]{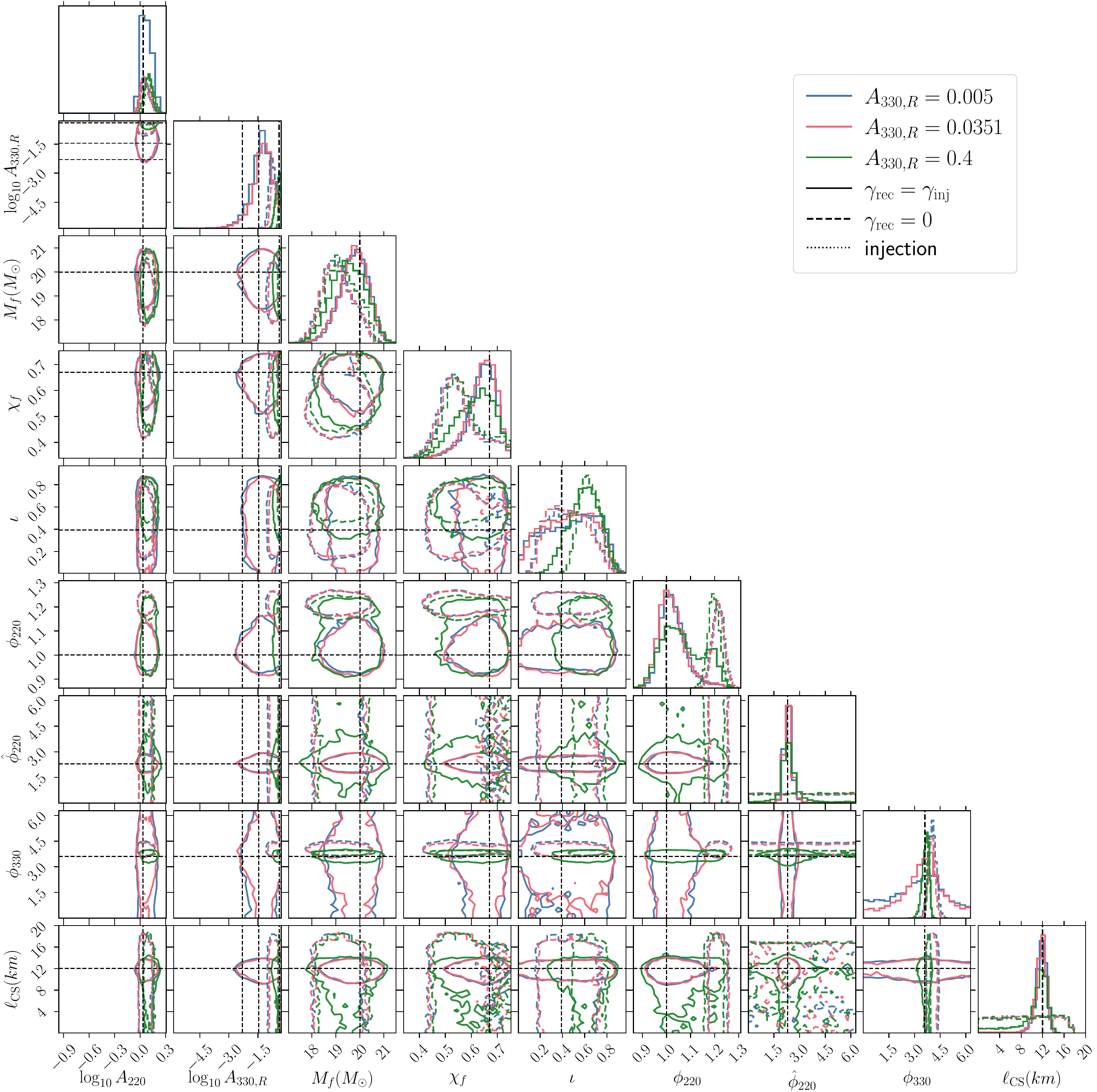}
  \caption{Same as Fig. \ref{fig:corner-esgb-phi0}, but for CS and $\gamma_{\rm inj}=10$.}
  \label{fig:corner-dcs-phi10}
\end{figure*}

In this subsection, we report additional results of the analyses performed. In Fig. \ref{fig:sampled_gamma_fig2}, we report the analog of the top-left panel of Fig. \ref{fig:alpha_0_3_cases} (i.e. the CS case), but including $\gamma$ in the sampling parameters. In Table~\ref{tab:BF_table_SNR_20_gamma} instead, we report the analog of Table~\ref{tab:BF_table_SNR_20} for the CS case, where $\gamma$ is sampled. Further, in Figs.~\ref{fig:corner-esgb-phi0}-\ref{fig:corner-dcs-phi10} we report the full corner plots related to Fig.~\ref{fig:330}, for the cases of $\gamma_{\rm inj}=0$ and $\gamma_{\rm inj}=10$.

\begin{table}[h!]
\centering
\begin{tabular}{|c|c|c|c|}
\hline
Theory & $\log_{10}\mathcal{B}(\gamma=1)$ & $\log_{10}\mathcal{B}(\gamma=5)$ & $\log_{10}\mathcal{B}(\gamma=10)$ \\
\hline
CS & 0.06 & -0.42  & -1.75 \\
\hline
\end{tabular}
\caption{Same as first row of Tab. \ref{tab:BF_table_SNR_20}, but for the analysis where $\gamma$ is sampled.}
\label{tab:BF_table_SNR_20_gamma}
\end{table}

\newpage

\bibliography{main}

\end{document}